\documentclass[aip,apl,reprint, superscriptaddress,amsmath,amssymb]{revtex4-1}

\usepackage{graphicx}
\usepackage{color}
\usepackage{units}
\begin{document}


\title{Narrow-band single photon emission at room temperature based on a single Nitrogen-vacancy center coupled to an all-fiber-cavity}



\author{Roland Albrecht}
\author{Alexander Bommer}
\affiliation{Universit\"at des Saarlandes, Fachrichtung 7.2 (Experimentalphysik), Campus E2.6, 66123 Saarbr\"ucken, Germany}
\author{Christoph Pauly}
\author{Frank M\"ucklich}
\affiliation{Universit\"at des Saarlandes, Fachrichtung 8.4 (Materialwissenschaft und Werkstofftechnik), Campus D3.3, 66123 Saarbr\"ucken, Germany}
\author{Andreas W. Schell}
\author{Philip Engel}
\affiliation{Humboldt-Universit\"at zu Berlin, Institut f\"ur Physik, AG Nanooptik, Newtonstra{\ss}e 15, 12489 Berlin, Germany}
\author{Tim Schr\"oder}
\affiliation{Department of Electrical Engineering and Computer Science, MIT, Cambridge, Massachusetts 02139, United States}
\affiliation{Humboldt-Universit\"at zu Berlin, Institut f\"ur Physik, AG Nanooptik, Newtonstra{\ss}e 15, 12489 Berlin, Germany}
\author{Oliver Benson}
\affiliation{Humboldt-Universit\"at zu Berlin, Institut f\"ur Physik, AG Nanooptik, Newtonstra{\ss}e 15, 12489 Berlin, Germany}
\author{Jakob Reichel}
\affiliation{Laboratoire Kastler Brossel, ENS/UPMC-Paris 6/CNRS, 24 rue Lhomond, 75005 Paris, France}
\author{Christoph Becher}
\affiliation{Universit\"at des Saarlandes, Fachrichtung 7.2 (Experimentalphysik), Campus E2.6, 66123 Saarbr\"ucken, Germany}
\email{c.becher@physik.uni-saarland.de}
\newcommand{\tc}{\textcolor{red}}

\date{\today}

\begin{abstract}
We report the realization of a device based on a single Nitrogen-vacancy (NV) center in diamond coupled to a fiber-cavity for use as single photon source (SPS). The device consists of two concave mirrors each directly fabricated on the facets of two optical fibers and a preselected nanodiamond containing a single NV center deposited onto one of these mirrors. Both, cavity in- and output are directly fiber-coupled and the emission wavelength is easily tunable by variation of the separation of the two mirrors with a piezo-electric crystal. By coupling to the cavity we achieve an increase of the spectral photon rate density by two orders of magnitude compared to free-space emission of the NV center. With this work we establish a simple all-fiber based SPS with promising prospects for the integration into photonic quantum networks.
\end{abstract}


\maketitle

Motivated by the prospects of quantum information science, manipulating and coupling of single quantum systems has inspired much research. In the past decade the Nitrogen-Vacancy (NV) center in diamond has been established as promising candidate for a solid state qubit due to its long spin coherence time~\cite{Jelezko04, Balasubramanian09} and as stable and bright single photon source (SPS) operating at room temperature.~\cite{Kurtsiefer00, Brouri00} For incorporation into quantum networks direct coupling of the photons to a single mode fiber is desired. Efforts have been made to deposit nanodiamonds (NDs) on photonic crystal fibers~\cite{Schroeder11.2} or tapered optical fibers.~\cite{Schroeder12, Liebermeister14} A drawback for many applications is the NV center's spectrally broad free-space emission (FWHM $\approx \unit[100]{nm}$). However, this can be overcome via coupling to an optical cavity: even for a NV center at room temperature cavity-enhanced narrow-band emission beyond simple spectral and spatial filtering can be observed.~\cite{Albrecht13, Kaupp13} 

Here we present an approach combining the cavity-enhanced emission with a simple, all-fiber based device, where excitation and collection are both fiber-coupled. We use a Fabry-Perot cavity consisting of two concave mirrors that are directly fabricated on the facets of optical fibers. Deterministic transfer of a ND onto one of these mirrors allows to couple a pre-selected single NV center to the cavity. By employing a flexure mount we realize a small, robust and compact device, where no further alignment of the cavity is necessary. The only remaining mechanical degree-of-freedom is the mirror separation which is controlled by a piezo-electric crystal enabling tuning of the cavity resonance frequency. The miniaturized footprint of fiber cavities (fiber diameter $D = \unit[125]{\mu m}$) ensures the scalability of the system~\cite{Derntl13} providing an easy integration into photonic networks. This very versatile device features a large tuning range which is only limited by the mirror coating and works in principle both at room and cryogenic temperatures.

\begin{figure}[t]
\centering\includegraphics[width=\columnwidth]{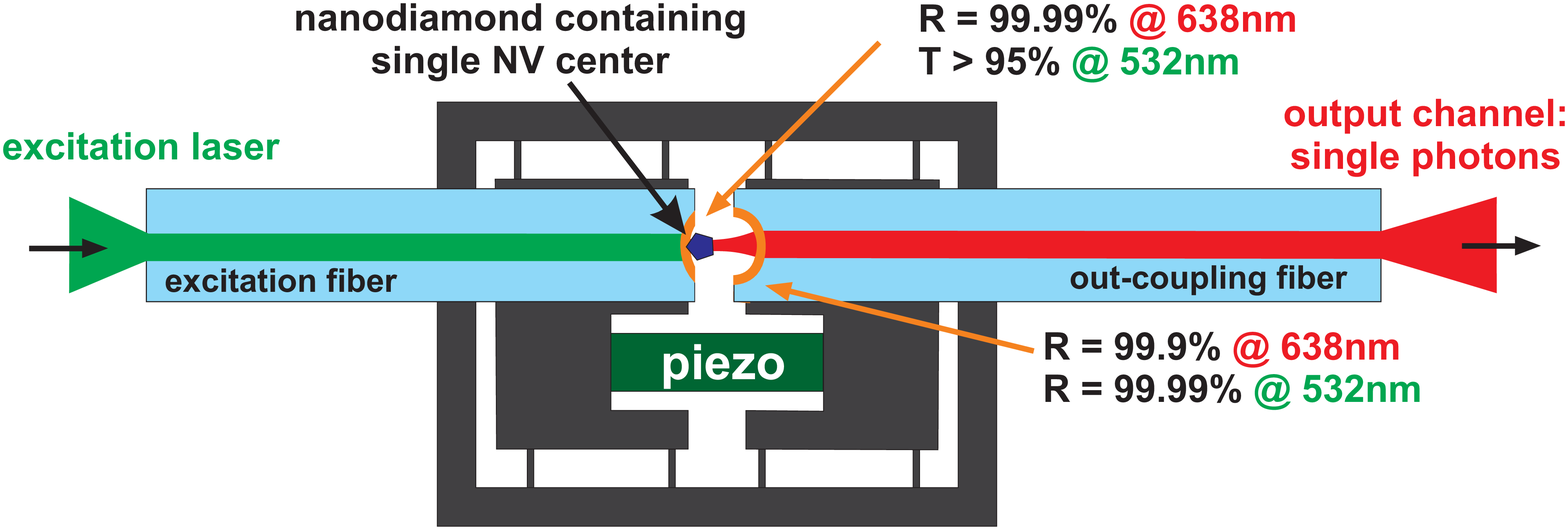}
\caption{Schematic of the device: An all-fiber based Fabry-Perot cavity consisting of two concave mirrors fabricated on the facets of optical fibers. The fibers are fixed to a flexure mount and the mirror separation is controlled via a  piezo-electric crystal. Onto one of the two fiber facets a pre-selected ND containing a single NV center is deposited. The NV center is excited directly through this fiber (excitation fiber) with a laser at \unit[532]{nm}. The mirror coatings are designed such that at the zero-phonon-line wavelength the dominant output channel is the second fiber (out-coupling fiber). Furthermore, the excitation (out-coupling) fiber mirror has a high transmission (reflectivity) for the excitation wavelength. The excitation (out-coupling) fiber has been FIB (laser) machined.}
\label{Fig1}
\end{figure}

The schematic of the device is illustrated in Fig.~\ref{Fig1}. It consists of two opposing optical fibers. Prior to the deposition of a highly reflective mirror coating onto the facets a concave imprint on each fiber was fabricated. For the proof-of-principle experiments shown in this letter we have employed two different fabrication methods. One of the fibers has been laser-machined using a CO$_2$ laser focused on the facet for a well defined exposure time.~\cite{Hunger12} This resulted in a concave structure with a radius of curvature $R = \unit[71.6]{\mu m}$, diameter $D = \unit[20.6]{\mu m}$ and a total depth (maximal vertical distance from highest point on facet to center of imprint) $t = \unit[1.9]{\mu m}$. The melting and re-crystallization of the glass surface during this process leads to a rms surface roughness $\sigma_\mathrm{rms}$ of a few {\AA}ngstr{\"o}m.~\cite{Hunger12} Such a surface roughness in principle enables cavities with a finesse up to $\mathcal{F} = 10^6$.~\cite{Muller10}

\begin{figure}[t]
\centering\includegraphics[width=\columnwidth]{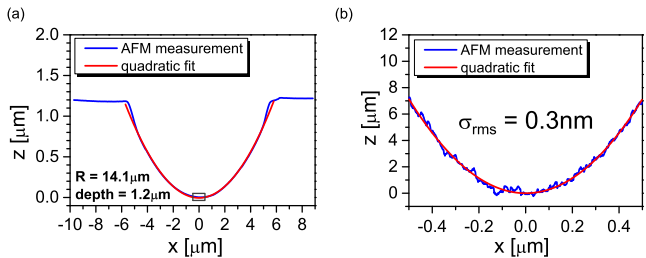}
\caption{(a) Cross-section through the profile of the FIB-machined imprint measured with an atomic force microscope and fitted with a quadratic function. The parabolic structure is $\unit[1.2]{\mu m}$ deep and has a radius of curvature of $R = \unit[14.1]{\mu m}$. (b) The center region of (a) revealing the smooth surface with a roughness of $\sigma_\mathrm{rms} = \unit[0.3]{nm}$.}
\label{Fig2}
\end{figure}

For the second fiber we have used focused ion beam (FIB) milling to fabricate a parabolic imprint~\cite{Dolan10} with radius of curvature $R=\unit[14.1]{\mu m}$, diameter $D = \unit[11.1]{\mu m}$ and depth $t=\unit[1.2]{\mu m}$. The cross-section through the center of the profile measured with an AFM is shown in Fig.~\ref{Fig2}. The radius of curvature has been determined via a quadratic fit of the profile. As for the laser machined fibers we obtain a very smooth surface with a rms surface roughness of $\sigma_\mathrm{rms} = \unit[0.3]{nm}$ determined on the $\unit[1.25\times1.25]{\mu m^2}$ profile's center area using a paraboloid as reference surface.

\begin{figure}[t]
\centering\includegraphics[width=\columnwidth]{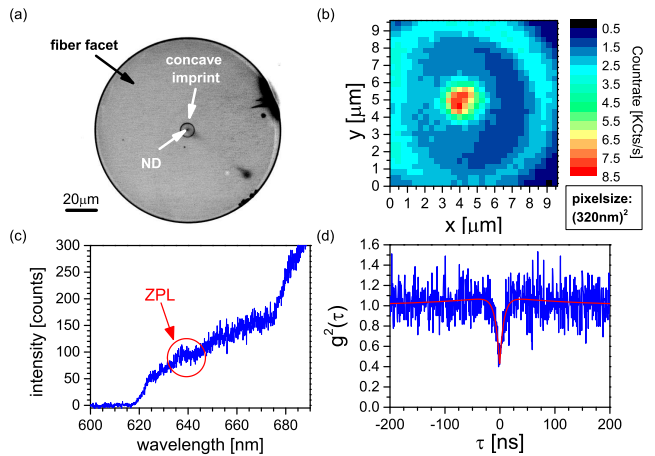}
\caption{(a) Microscope image of a fiber facet (diameter $\unit[125]{\mu m}$) with a concave parabolic imprint fabricated via FIB milling. In its center a ND containing a single NV center has been deposited. (b) Fluorescence microscope image of a structured fiber: The excitation laser is coupled into the fiber from the non-structured end. A microscope objective in front of the fiber mirror collects the fluorescence  which is then focused onto a $\unit[50]{\mu m}$ core of a multimode fiber with a $f = \unit[50]{mm}$ achromatic doublet. Recording of the fluorescence intensity in the spectral window $\unit[633-647]{nm}$ during a lateral scan of the fiber yields the fluorescence microscope image. (c) Emission spectra from the ND at the center of the imprint excited through the fiber. At $\approx \unit[640]{nm}$ the ZPL of the NV center is visible. Above $\unit[680]{nm}$ the transmission of the mirror coating strongly increases such that the spectrum is dominated by background fluorescence generated within the excitation fiber. (d) Measured intensity correlation function of the emission from the ND in the spectral window $\unit[633-647]{nm}$ revealing emission of single photons from the single NV center with $g^{(2)}(0) = 0.40$.}
\label{Fig3}
\end{figure}

Onto one of the fiber mirrors a ND containing a single NV center is deposited. The NV center is directly excited through the fiber by a laser at a wavelength $\lambda_\mathrm{exc} = \unit[532]{nm}$ coupled into the fiber core. Henceforth we call this fiber the excitation fiber. The NV center has been pre-selected: first, the potential NDs (Microdiamant, MSY GAF 0-0.05) are spin coated on a cover slip. A well-suited ND containing a single NV center is identified, which is subsequently transferred onto the fiber mirror using an atomic force microscope based pick-and-place technique.~\cite{Schell11.2} After the ND is placed on the facet, it is moved to its final position at the center of the concave imprint using the atomic force microscope in contact mode. Figure~\ref{Fig3}(a) shows a microscope image of the transferred ND at the center of the fiber facet.

We use a Fabry-Perot cavity with asymmetric reflectivities for the design wavelength, the NV center's zero-phonon-line (ZPL) wavelength ($\lambda_\mathrm{ZPL} = \unit[638]{nm}$), such that the second fiber, the out-coupling fiber, acts as dominant output channel. The coating on the excitation fiber, which is a pure silica core, single mode fiber (Nufern S-460HP), is designed such that the field intensity at the mirror surface, where the ND is positioned, is maximal and it has a high transmission for the excitation wavelength. The pure silica core ensures low background fluorescence generated within the excitation fiber. As out-coupling fiber we use a gold-coated single mode fiber (Fiberguide Industries ASI4.3/125/155G). The background fluorescence is prevented by employing a mirror coating with a high reflectivity for the excitation wavelength. We here aim for a moderate finesse of approximately $\mathcal{F}\approx 4000$ at the ZPL wavelength.

We first verify that direct excitation of the NV center through the fiber is possible. The excitation laser is coupled into the fiber from the non-structured end and a microscope objective in front of the fiber mirror collects the emitted fluorescence from a localized spot on the fiber facet. To suppress the excitation laser a dielectric bandpass filter with transmission edge at $\unit[620]{nm}$ is employed. A lateral scan of the facet [cf. Fig~\ref{Fig3}(b)] reveals the edges of the concave imprint and a bright spot in its center, where the NV is located. In the emission spectra [cf. Fig~\ref{Fig3}(c)] the ZPL of the NV center is visible. The intensity correlation function $g^{(2)}(\tau)$ shows a clear anti-bunching at zero time-delay below 0.5 [cf. Fig~\ref{Fig3}(d)]. Thus we can directly excite the single NV center through the fiber while maintaining the single photon statistics of its emission. We assume an effective 3-level model to describe the internal dynamics and therefore fit the intensity correlation function with 
\begin{equation}
g^{(2)}(\tau) = g_N + (1-g_N)[1 - (1+a)e^{-|\tau|/\tau_1} +  a e^{-|\tau|/\tau_2}]
\end{equation} 
where $\tau_1$ ($\tau_2$) describe the shape of the anti-bunching (bunching) and $a$ determines the strength of bunching. The noise contribution to the $g^{(2)}$ is included with $g_N$. The signal to noise ratio (SNR) amounts to $\mathrm{SNR} = \sqrt{1-g_N}/(1-\sqrt{1-g_N}) = 3.4$. Measuring the number of photons emitted in the spectral window of \unit[633-647]{nm} as function of excitation power reveals a saturation count rate of $\mathrm{CR}_\infty = \unit[41.8]{KCts/s}$ and a saturation power of $P_\mathrm{sat} = \unit[12.6]{mW}$ (i.e the power out-coupled from the fiber). Taking into account the mode field diameter of the fiber of $w_\mathrm{MFD} = \unit[4.3]{\mu m}$ the saturation intensity is $I_\mathrm{sat} = \unit[8.7\times 10^4]{W/cm^2}$.

\begin{figure}[t]
\centering\includegraphics[width=\columnwidth]{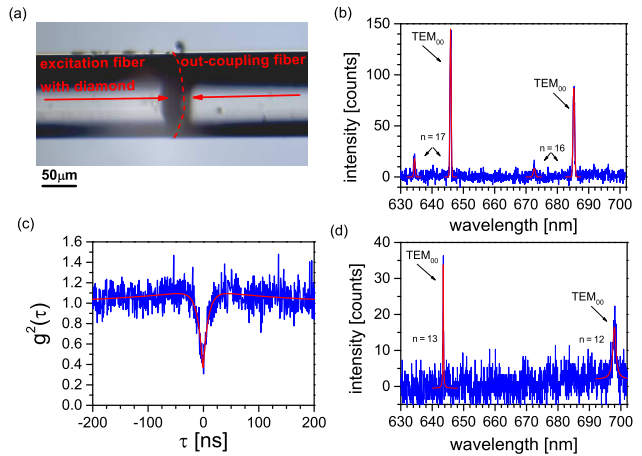}
\caption{(a) Microscope image of an all-fiber based Fabry-Perot cavity. As there is no more visible gap between the two fibers, the separation of the fibers is sketched with the red dashed line. (b) Emission spectrum of the Fabry-Perot cavity with effective length $l_\mathrm{cav} = \unit[5.6]{\mu m}$. There are two fundamental Gaussians TEM$_{00}$ modes visible with longitudinal mode numbers $n = 16$ and $n = 17$ ($l = n \times \lambda_\mathrm{cav}/2$). We do not observe any polarization splitting of the fundamental modes within the cavity bandwidth. Besides emission into the fundamental mode, there is also weak emission into higher order transverse modes. (c) Intensity correlation measurement $g^{(2)}(\tau)$ of the cavity emission revealing emission of single photons with $g^{(2)}(0) = 0.34$. (d) Cavity emission spectrum of cavity with effective length $l_\mathrm{cav} = \unit[4.1]{\mu m}$. This is the smallest cavity length reachable with the given concave profiles of the two fibers.}
\label{Fig4}
\end{figure}

With this fiber we now build a Fabry-Perot cavity as shown in Fig~\ref{Fig4}(a). In contrast to the broad emission spectrum for free-space emission we now observe only emission into narrow cavity lines. Figure~\ref{Fig4}(b) shows the emission spectrum of the output from the out-coupling fiber where the effective cavity length has been set to $l_\mathrm{cav} = \unit[5.6]{\mu m}$. There are two longitudinal modes visible in the spectral region from $\unit[630-700]{nm}$, at $\lambda = \unit[645.9]{nm}$ and at $\lambda = \unit[685.3]{nm}$. The cavity length has been determined from the direct visible free spectral range $\nu_\mathrm{FSR} = \unit[26.7]{THz}$. The waist of the Gaussian mode is determined by the two radii of curvature and the length as $w_0 = \unit[1.2]{\mu m}$. The mode volume calculates to $V_\mathrm{mod} = \pi/4 \times w_0^2l = \unit[6.1]{\mu m^3} = \unit[22.7]{\lambda_\mathrm{cav}^3}$. From a saturation measurement we determine a maximal count rate $CR_\infty = \unit[2800]{Counts/s}$ emitted into the mode at $\lambda = \unit[645.9]{nm}$. Monitoring the reflection and transmission of an external cavity diode laser at a wavelength of approximately \unit[640]{nm} while scanning the cavity length yields a finesse of $\mathcal{F} = 3600$. This is the same value as we measure for a control cavity without a ND. Thus, the single, deterministically placed ND does not induce any significant scattering losses at this level of finesse. The spectral width of the resonance amounts to $\Delta\nu = \unit[7.6]{GHz}$. Hence, the spectral photon rate density for the cavity-coupled emission computes to $\unit[380]{photons/(s~GHz)}$. This is two orders of magnitude larger compared to the free-space emission where we obtain a value of $\unit[41.8]{KCts/s}/\unit[10.2]{THz} = \unit[4.1]{photons/(s~GHz)}$. This clearly proves that we here observe cavity-enhanced emission. Using a theoretical model describing the cavity coupling~\cite{Albrecht13, Auffeves09} we expect a saturation count rate of $CR_{\infty\mathrm{,~theo}} = \unit[5700]{Counts/s}$ into the fundamental mode at a wavelength of $\unit[645.9]{nm}$ for an ideally oriented NV center (assuming dipoles oriented perpendicular to optical axis), where 1.8\% of the total NV center's emission is emitted via the cavity. We observe by a factor of 2 less photons because of non-perfect orientation of the dipole moment and a non-unity quantum-efficiency of the NV center.

Measurement of the intensity correlation function $g^{(2)}$ of the cavity output confirms single photon emission as shown in Fig.~\ref{Fig4}(c). Because only emission from the NV center is cavity-enhanced we observe a lower $g^{(2)}(\tau=0) = 0.34$ corresponding to an increased $\mathrm{SNR = 4.4}$ compared to emission into free-space. To increase the cavity coupling we reduce the cavity length. The emission spectrum for the smallest reachable cavity length is displayed in Fig.~\ref{Fig4}(d). At this point the two fibers are in contact. From the free spectral range $\nu_\mathrm{FSR} = \unit[36.3]{THz}$ we deduce a cavity length of $l_\mathrm{cav} = \unit[4.1]{\mu m}$. Note that the smallest reachable effective cavity length depends not only on the depths of the imprints ($t_1=\unit[1.2]{\mu m}$ and $t_2=\unit[1.9]{\mu m}$ for the two fibers used here), but additionally on the penetration depth of the electric field into the mirror coatings. The mode volume is reduced to $V_\mathrm{mod} = \unit[4.1]{\mu m^3} = \unit[15.5]{\lambda_\mathrm{cav}^3}$ and we observe a photon rate increased by a factor of approximately 1.5. As for the smaller cavity the linewidth is increased to $\Delta\nu = \unit[10.1]{GHz}$ the spectral photon rate density is only slightly increased to $\unit[420]{photons/(s~GHz)}$.

To further increase the efficiency we need an even smaller mode volume. A cavity with two FIB machined fibers with both mirror radii of curvature of $R=\unit[25]{\mu m}$ and an effective cavity length $l_\mathrm{eff} = \unit[2.23]{\mu m} = 3.5\times\lambda_\mathrm{NV}$ is realizable. This leads to a mode volume of $V_\mathrm{mod} = \unit[1.84]{\mu m^3}$. Assuming a realistic finesse~\cite{Hunger12} of $\mathcal{F} = 5\times 10^4$ we use our theoretical model~\cite{Albrecht13} to predict an efficiency of the SPS of 6.9\% with a spectral bandwidth of $\Delta\nu = \unit[1.3]{GHz}$ while the cavity is in resonance with the ZPL. As efficiency we here define the portion of the total NV emission, that is channeled into the cavity mode and dominantly out-coupled through the fiber. The spectral bandwidth is given by the linewidth of the cavity mode. At room temperature this efficiency of the SPS is limited because of the low Debye-Waller factor of the NV center ($\approx 0.03$) and the ZPL linewidth. Cooling of the cavity and NV center to cryogenic temperature reduces the ZPL linewidth and increases the efficiency of the SPS to 90\% assuming a ZPL linewidth at cryogenic temperatures of $\Delta\nu_\mathrm{ZPL} = \unit[10]{GHz}$.

A promising alternative color center in diamond is the SiV center~\cite{Neu11}. It shows bright emission at $\unit[738]{nm}$ with photon-rates $> \unit[5\times10^6]{photons/s}$, where 80\% of the emission is into the ZPL. Its drawback is a low quantum efficiency of approximately 5\%.~\cite{Neu12} We can estimate the performance of a fiber-cavity based SPS with a single SiV center: Assuming the same cavity parameters as above (both radii of curvature of $R=\unit[25]{\mu m}$, effective cavity length $l_\mathrm{eff} = \unit[2.22]{\mu m} = 3\times\lambda_\mathrm{SiV}$ and finesse $\mathcal{F} = 5\times 10^4$) and taking into account the typical SiV lifetime of 1ns, the ZPL's room temperature width $\Delta\nu_\mathrm{SiV, ZPL} = \unit[550]{GHz}$, the low quantum efficiency and the SiV's Debye-Waller factor of 0.8, we expect for the all-fiber-cavity based SPS an efficiency of 30\% where the SiV decays with a rate of $\unit[430]{MHz}$ into the cavity mode. The effective photon count rate is then given by this rate times the out-coupling efficiency of the cavity ($\approx 50\%$) and the detection efficiency. The SPS operates at room temperature and emits single photons within a spectral bandwidth of $\unit[1.3]{GHz}$.

In conclusion, we here have introduced a versatile compact directly fiber-coupled SPS. The deterministic incorporation of pre-selected ND containing a single NV center into a fiber-based Fabry-Perot cavity makes the system reproducible and easily scalable. With cavity-enhanced narrow-band single photon emission and both excitation and output fiber-coupled the SPS is highly suitable for integration into photonic quantum networks. 

\begin{acknowledgments}
We acknowledge partial financial support by the European Community's Seventh Framework Programme (FP7/2007-2013) under Grant Agreement No618078 (WASPS) and No611143 (DIADEMS), by the EU funding for the project AME-Lab (European Regional Development Fund C/4-EFRE 13/2009/Br) for the FIB/SEM and by the Deutsche Forschungsgemeinschaft (DFG, FOR1493).
\end{acknowledgments}




%

\end{document}